\documentclass[usenatbib,useAMS]{mn2e}

\usepackage{natbib}
\usepackage{amsmath}
\usepackage{graphicx}
\usepackage{color}
\usepackage{mathrsfs}
\usepackage{epsfig}
\allowdisplaybreaks[1]

\newcommand{\msun}{{\rm M}_\odot}

\newcommand{\zsun}{Z_\odot}

\newcommand{\mpc}{{\rm Mpc}}
\newcommand{\gpc}{{\rm Gpc}}

\newcommand{\beq}{\begin{equation}}
\newcommand{\eeq}{\end{equation}}

\defcitealias{K14}{K14}
\defcitealias{K16}{K16}
\defcitealias{Visbal_2015}{VHB15}

\title[Gravitational wave background from PopIII BHs]
{Gravitational wave background from Population III binary black holes
consistent with cosmic reionization}

\author[]{Kohei Inayoshi$^{1}$\thanks{E-mail: inayoshi@astro.columbia.edu},
Kazumi Kashiyama$^{2}$,
Eli Visbal$^{1}$,
Zolt\'an Haiman$^{1}$\\
$^{1}$Department of Astronomy, Columbia University, 550 W. 120th Street, New York, NY 10027, USA\\
$^{2}$Department of Astronomy; Theoretical Astrophysics Center; University of California, Berkeley, CA 94720, USA
}

\begin{document}
\maketitle
\label{firstpage}

\begin{abstract}
The recent discovery of the gravitational wave source GW150914 has revealed
a coalescing binary black hole (BBH) with
masses of $\sim 30~\msun$.  
Previous proposals for the origin of such a massive binary include Population III (PopIII) stars.
PopIII stars are efficient producers of BBHs and of a
gravitational wave background (GWB) in the $10-100$ Hz band, and also
of ionizing radiation in the early Universe.  
We quantify the relation between the amplitude of the GWB
($\Omega_{\rm gw}$) and the electron scattering optical depth
($\tau_{\rm e}$), produced by PopIII stars, assuming that 
$f_{\rm esc}\approx 10\%$ of their ionizing radiation escapes into the
intergalactic medium.  We find that PopIII stars would produce a
GWB that is detectable by the future O5 LIGO/Virgo if $\tau_{\rm e} \ga 0.07$,
consistent with the recent Planck measurement of $\tau_e=0.055 \pm 0.09$.
Moreover, the spectral index of the background from PopIII BBHs becomes 
as small as ${\rm d}\ln \Omega_{\rm gw}/{\rm d}\ln f\la 0.3$ at $f \ga 30$ Hz, 
which is significantly flatter than the value $\sim 2/3$ generically produced 
by lower-redshift and less-massive BBHs.  
A detection of the unique flattening at such low frequencies by
the O5 LIGO/Virgo will indicate the existence of a
high-chirp mass, high-redshift BBH population, which is consistent
with the PopIII origin.
A precise characterization of the spectral shape near $30-50$ Hz
by the Einstein Telescope could also constrain the PopIII initial
mass function and star formation rate.
\end{abstract}

\begin{keywords}
gravitational waves -- black hole physics -- stars: Population III
\vspace{-1\baselineskip}
\end{keywords}


\section{Introduction}
\label{sec:intro}

Advanced LIGO (AdLIGO) announced the first direct detection of
gravitational waves. The source, GW150914, is inferred to be a merging
binary black hole (BBH) with masses of $36^{+5}_{-4}~\msun$ and
$29^{+4}_{-4}~\msun$ at $z = 0.09_{-0.03}^{+0.04}$
\citep{Abbott_PRL_2016}.  The origin of such massive and compact BBHs
are of astrophysical interest, and several pathways have been proposed
for their formation. One channel is massive binary evolution in a
metal-poor environment~(\citealt{1984MNRAS.207..585B}; 
\citealt{Belczynski_2004,2006A&A...459.1001K}; 
\citealt{,Dominik_2012,K14,K16}, hereafter K14,
K16; \citealt{Belczynski_2016_b}), including rapid rotation and tides
\citep{Mandel_2016}, or assisted by accretion discs in active galactic
nuclei (\citealt{Bartos_2016,Stone_2016}).  Alternative formation
channels include stellar collisions in dense clusters
\cite[e.g.][]{PortegiesZwart_2000,OLeary_2016} or the collapse 
of rapidly-rotating massive
stars~\citep{Loeb_2016}.  Given the estimated BBH merger
rate of $\sim 2-400~{\rm Gpc}^{-3}~{\rm yr}^{-1}$
\citep{Abbott_2016_rate}, these scenarios may be distinguished in the
near future by their chirp mass distributions.

Another way to probe the star and BH formation history of the Universe
by GWs is via the stochastic background (GWB) from numerous unresolved BBH
mergers.  Here we focus on one promising source, binary Population~III
(hereafter PopIII) stars formed in the early Universe at $z \ga 6$.
PopIII stars are thought to be massive~\citep[][and references
  therein]{2011ARA&A..49..373B}, and their binary fraction is expected
to be at least as high as for present-day massive
stars~\citep[e.g.][]{2013MNRAS.433.1094S}.
Thus PopIII stars may form massive BBHs effectively\footnote{
For the purposes of this paper, "PopIII" refers
to any stellar population with unusually high typical mass, forming
predominantly at high redshift ($z>6$), and contributing to reionization
at these redshifts. The IMF as a function of metallicity is poorly
understood, and these properties may also be satisfied for an extreme
"PopII" stellar population, enriched to low, but non-zero metalicity
(e.g. $Z<0.1~\zsun$; see \citealt{Kowalska_2015,Belczynski_2016_b}).}.
The vast majority of these BBHs merge at high redshifts and contribute to the
GWB \citep{Kowalska_2012}.  
A small minority merge after a delay comparable to the Hubble
time, inside the detection horizon of AdLIGO.  Recent PopIII binary
population synthesis calculations predicted the event rate at $z\simeq
0$ to be $\sim 1-100~{\rm Gpc^{-3}}~{\rm yr}^{-1}$, and the chirp mass
distribution to peak at $\sim 30~\msun$ (\citetalias{K14,K16}), in
near-perfect agreement with GW150914.

PopIII stars are also efficient producers of ionizing radiation in
the early Universe.  Recently, the optical depth of the universe to
electron scattering was inferred from the cosmic microwave background
(CMB) anisotropies by the {\it Planck} satellite to be $\tau_{\rm e} =
0.066 \pm 0.016$ \citep{Planck_2015}.  This value is lower than
previous estimates from {\it WMAP} \citep{Komatsu_2011}, placing tight
constraints on the formation history of PopIII stars
\citep[][hereafter VHB15]{Visbal_2015}.

In this paper, we show that a PopIII BBH population that is
consistent with the {\em Planck} optical depth measurement produces a
GWB dominating over other BBH populations at $10-100$ Hz.
As a result, the stochastic GWB may be detectable sooner than
previously expected.  We also find that the spectral index of the GWB
due to PopIII BBHs becomes significantly flatter at $f\ga 30$ Hz than the value
${\rm d}\ln \Omega_{\rm gw}/{\rm d}\ln f \approx 2/3$ generically produced by
the circular, GW-driven inspiral of less massive and/or lower-$z$ BBHs
\citep{Phinney_2001}.  This flattening can not be mimicked by
eccentric orbits or dissipative processes, and could be detected by
the observing run O5 by AdLIGO/Virgo.  Such a detection would provide
robust evidence for a high-chirp-mass, high-redshift BBH population,
and would yield information about the initial mass function (IMF) and
star formation rate (SFR) of PopIII stars, and about cosmic reionization.
\footnote{As this paper was being completed, we became
    aware of a recent preprint \citep{Hartwig_2016}, which presents a
    more conservative estimate for the PopIII SFR and the
    corresponding GW emission.  We here highlight the possibility of a
    higher PopIII SFR and GWB, which are still consistent with the
    {\it Planck} results.}


\section{Massive binary formation rate at high-redshifts}
\label{sec:BBH}

\subsection{PopIII stars $(Z=0)$}

PopIII stars are the first generation of stars in the Universe,
beginning to form from pristine gas in mini-halos with masses of
$10^{5-7}~\msun$ at redshifts $z\sim 20-30$.  These stars are likely
to be massive, because of inefficient cooling via molecular hydrogen
(H$_2$), resulting in a top-heavy IMF covering the mass range
$\sim10-300~\msun$ \citep[e.g.][]{2014ApJ...781...60H}.  Massive
PopIII stars produce a strong Lyman-Werner (LW) radiation background,
which dissociates H$_2$ in mini-halos and suppresses subsequent star
formation \citep[e.g.][]{1997ApJ...476..458H}.  Including this
self-regulation, the history of early star-formation has been
investigated in semi-analytical studies \citep[e.g.][]{Haiman_2000,
  SobacchiMesinger_2013} and in cosmological simulations
\citep[e.g.][]{Ricotti_2002, Ahn_2012}.

The {\rm Planck} measurement of the optical depth from the CMB
($\tau_{\rm e}=0.066\pm \Delta \tau_{\rm e}$, where $\Delta \tau_{\rm
  e}=0.016$ is the $1\sigma$ error; \citealt{Planck_2015}) tightly
constrains the PopIII star formation history.
\citetalias{Visbal_2015} estimated an upper limit on the total mass density
of PopIII
stars as $\rho_{\rm \ast III}\sim 10^5~\msun~\mpc^{-3}$, consistent
with the Planck result.  They assumed a top-heavy IMF, with PopIII
stars as massive as $\sim 200~\msun$, for which the number of
H-ionizing and LW photons per stellar baryon are both $\eta_{\rm
  ion}\simeq \eta_{\rm LW}\simeq 8\times 10^4$; the escape fraction of
ionizing photons from mini-halos was assumed to be $f_{\rm
  esc,m}=0.5$.

We explore the dependence of $\rho_{\rm \ast III}$ on $\eta_{\rm
  ion(LW)}$, $f_{\rm esc,m}$ and $\tau_{\rm e}$.  We consider two
cases for a flat and a Salpeter IMF with $10-100~\msun$.
The ionizing photon number per baryon is $\eta_{\rm
  ion}=7.1~(5.1)\times 10^4$ for the flat (Salpeter) IMF
\citep{2002A&A...382...28S}.  Note that the value of $\eta_{\rm ion}$
is almost constant for a higher maximum mass of the IMF with $M_{\rm
  max}>100~\msun$.  We assume $\eta_{\rm ion}=\eta_{\rm LW}$, which is
a good approximation for these IMFs.  We consider two values of the
escape fraction, $f_{\rm esc,m}=0.1$ and $0.5$.  The higher value
corresponds to mini-halos with $10^{5-6}~\msun$
\citep{2004ApJ...613..631K}, while the lower value
corresponds to more massive halos with $\ga {\rm a~few}\times 10^7~\msun$
\citep{Wise_2014}.  In fact, the typical mass of halos forming PopIII
stars is $\simeq 10^{7-8}~\msun$ at $z\simeq 10$, because once a LW
background develops it disables H$_2$ cooling in smaller halos.  We
thus adopt $f_{\rm esc,m}=0.1$ as our fiducial model.

\begin{figure}
\begin{center}
\includegraphics[width=75mm]
{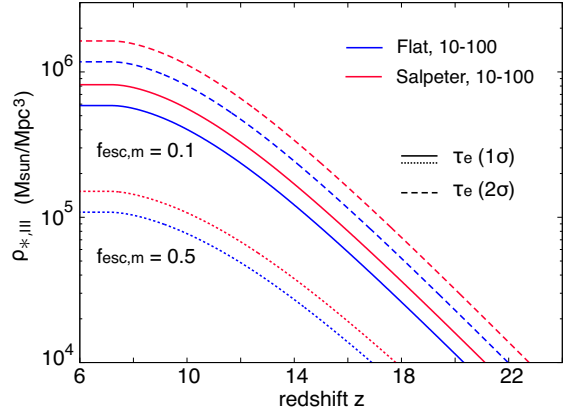}
\caption{Cumulative (comoving) mass density of PopIII stars
  consistent with the {\it Planck} optical depth measurement from the
  CMB for two different IMFs: flat (blue) and Salpeter (red) with
  $10-100~\msun$.  The optical depth is set to $\tau_{\rm
    e}=0.066+0.016$ ($1\sigma$; solid and dotted) and $0.066+0.032$
  ($2\sigma$; dashed), respectively.  Two cases with $f_{\rm
    esc,m}=0.1$ (solid and dashed) and $0.5$ (dotted) are shown.  At
  $z\la 8$, PopIII star formation is shut down 
  and the cumulative mass density saturates
  (see Eq. \ref{eq:rho_max}).  }
\label{fig:rho_pop3}
\end{center}
\end{figure}

Fig. \ref{fig:rho_pop3} shows the cumulative (comoving) mass density
of PopIII stars for different IMFs and escape fractions.  
All other model ingredients, apart from an overall normalization of
the SFR, follow the fiducial model of \citetalias{Visbal_2015}.
The normalization is chosen to produce an optical depth of $\tau_{\rm
  e}=0.066+\Delta \tau_{\rm e}$ (solid and dotted curves) and
$0.066+2\Delta \tau_{\rm e}$ (dashed curve).  In all cases, the mass
density of PopIII stars is saturated at $z\la 8$, where PopIII star
formation is shut down because of LW feedback, metal enrichment and reionization.
The saturated value can be given approximately by
\begin{align}
\rho_{\rm \ast,III}&\simeq 8.2\times 10^5~\msun~\mpc^{-3}\nonumber\\
&\times
\left(\frac{\eta_{\rm ion}}{5\times 10^4}\right)^{-1}
\left(\frac{f_{\rm esc,m}}{0.1}\right)^{-1}
\left(\frac{\tau_{\rm e}-0.066}{\Delta \tau_{\rm e}}\right),
\label{eq:rho_max}
\end{align}
for the range we consider here.
Note that \citetalias{K14} estimated the GW event rate from PopIII
binaries adopting a higher comoving mass density of $\rho_{\rm
  \ast,III}\simeq 2\times 10^6~\msun~{\rm
  Mpc}^{-3}$~\citep{DeSouza_2011}.

Next, we estimate the number density of massive PopIII binary stars
which potentially evolve into PopIII BBHs.  The distribution function
of the mass ratio $q\equiv M_{2}/M_{1}$ of binary stars is assumed to
be $\Phi(q)={\rm const}$, where $M_{1(2)}$ is the mass of the primary
(secondary) star.  This distribution is consistent with the slope 
${\rm d}\ln \Phi/{\rm d}\ln q = -0.1\pm0.6$,
observed for present-day massive binary stars \citep{Sana_2012}.
The mass ratio range for a fixed $M_1$ is
$q_{\rm min}(=M_{\rm min}/M_1)\leq q \leq q_{\rm max}(=M_{\rm max}/M_1)$.  
Thus, the number fraction of
stars which form massive binaries with 
$M_{1(2)}\geq M_{\rm crit}=25~\msun$, 
above which a star is assumed to collapse into a BH
\citep{2002ApJ...567..532H}, is estimated 
for a given IMF as
\begin{align}
f_{\rm MB} 
&= \frac{\int _{M_{\rm crit}}^{M_{\rm max}}dM_1\int_{q_{\rm crit}}^{q_{\rm max}}dq~\Phi(q)~\dfrac{dN}{dM_1}}
{\int _{M_{\rm min}}^{M_{\rm max}}dM_1\int_{q_{\rm min}}^{q_{\rm max}}dq~\Phi(q)~\dfrac{dN}{dM_1}},\nonumber\\[2pt]
=&
\frac{M_{\rm max}-M_{\rm crit}}{M_{\rm max}-M_{\rm min}}
\begin{cases}
\frac{\ln(M_{\rm max}/M_{\rm crit})}{\ln(M_{\rm max}/M_{\rm min})}~~~{\rm for~\alpha =0},\\[4pt]
\frac{M_{\rm crit}^{-\alpha}-M_{\rm max}^{-\alpha}}{M_{\rm min}^{-\alpha}-M_{\rm max}^{-\alpha}}~~~~~{\rm for~\alpha \neq 0},\\
\end{cases}
\label{eq:f_MB}
\end{align}
where $q_{\rm crit}=M_{\rm crit}/M_1$.  
For the flat ($\alpha=0$) and
Salpeter ($\alpha =2.35$) IMF with $10-100~\msun$, we estimate $f_{\rm
  MB}\simeq 9.3\times 10^{-2}$ and $0.5$, respectively.  From
Eqs. (\ref{eq:rho_max}) and (\ref{eq:f_MB}), the number density of the
massive binaries is estimated as
\begin{align}
N_{\rm MB,III}&=\frac{\rho_{\rm \ast,III}}{\langle M_\ast\rangle}\frac{f_{\rm bin}}{1+f_{\rm bin}}f_{\rm MB}\nonumber\\[0pt]
&=1.6\times 10^3~\mpc^{-3}~
\left(\frac{\eta_{\rm ion}}{50000}\right)^{-1}
\left(\frac{f_{\rm esc,m}}{0.1}\right)^{-1}\nonumber\\[0pt]
&\times \left(\frac{\langle M_\ast\rangle}{20~\msun}\right)^{-1}
\left(\frac{f_{\rm bin}/0.7}{1+f_{\rm bin}/0.7}\right)
\left(\frac{f_{\rm MB}}{0.1}\right),
\end{align}
where $\langle M_\ast\rangle$ is the average mass of single stars,
$f_{\rm bin}\equiv N_{\rm binary}/N_{\rm single}$ is the binary
fraction, and the fiducial $f_{\rm bin}=0.7$ is based on the binary
fraction $0.69\pm0.09$ measured for present-day massive stars
\citep{Sana_2012}.  Thus, the number density of PopIII BBHs is
estimated as $N_{\rm MB,III}\simeq 2.7~(1.4)\times 10^3~\mpc^{-3}$ for
the flat (Salpeter) IMF.

Fig. \ref{fig:MB_ion} shows the number of massive binaries per
ionizing photon, a quantity directly connecting the GWB to
reionization, from PopIII stars for the flat (blue) and Salpeter (red)
IMF with $5~\msun\leq M_{\rm min}\leq 25~\msun$.  
Note that stars with $M>25~\msun$ contribute both BBHs and ionizing
photons, while stars in the range $5-25~\msun$ add significant
ionizing photons without any additional BBHs.  As long as PopIII stars
are massive, with $M_{\rm min}\ga 10~\msun$, our results are
relatively insensitive to the slope of the IMF (less than a factor of
two).  However, if $M_{\rm min}\la 10~\msun$, normalizing the ionizing
emissivity to match the {\it Planck} optical depth implies fewer
massive binaries, and the precise value also depends more strongly on
the IMF slope.

\subsection{PopII stars $(Z\sim 0.1~\zsun)$}

Population II (PopII) stars are formed in metal-enriched gas clouds once supernovae of
massive PopIII stars have produced heavy elements.  Because metal/dust
cooling is more efficient than H$_2$ cooling, the IMF of PopII stars
is expected to be less
top-heavy~\citep[e.g.][]{2005ApJ...626..627O}.  We
assume that once metal enrichment has occurred, PopII stars form with
$\simeq 0.1~\zsun$, although the precise value of this metallicity does
not significantly affect our results.

\begin{figure}
\begin{center}
\includegraphics[width=75mm]
{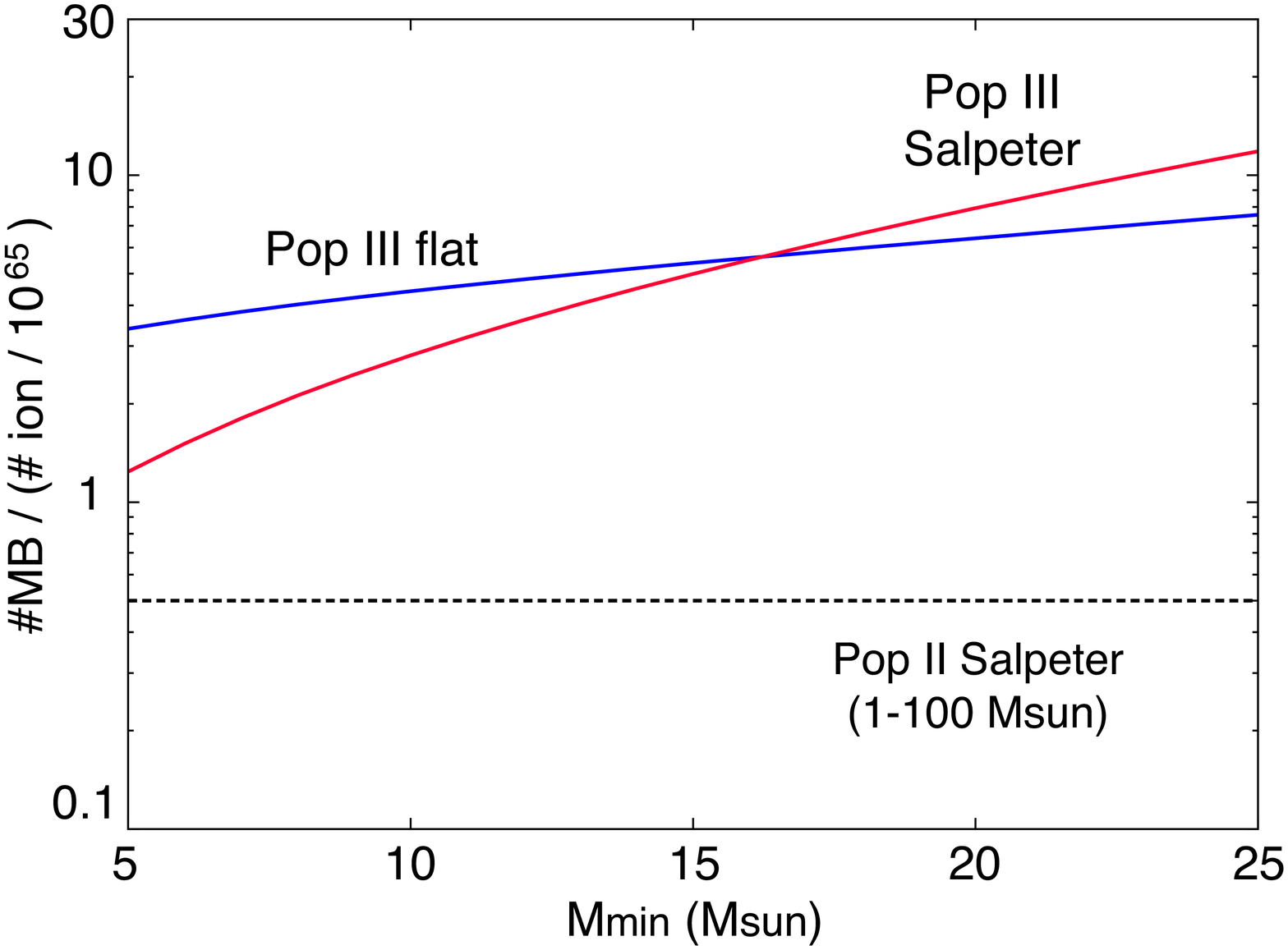}
\caption{The ratio of the number of massive BH binaries to the number
  of ionizing photons for a flat (blue) and Salpeter (red) PopIII IMF
  with $5~\msun\leq M_{\rm min}\leq 25~\msun$ and $M_{\rm
    max}=100~\msun$.  For reference, the ratio for a Salpeter IMF of
  PopII stars with $1~\msun \leq M \leq 100~\msun$ is shown (black
  dashed).  All curves assume a stellar binary fraction of $f_{\rm
    bin}=0.7$.  PopIII stars with $M_{\rm min}\ga 10~\msun$ produce an
  $\sim$ order of magnitude more BH binaries.  }
\label{fig:MB_ion}
\end{center}
\end{figure}

Observations of high-redshift galaxies provide estimates of the
stellar mass density
\citep[e.g.][]{2008ApJ...675..234P},
which are consistent with cosmological simulations
\citep[e.g.][]{2013MNRAS.436.3031V}.  At $z\sim 3$, the (comoving)
stellar mass density is $\rho_\ast \simeq 3\times
10^7~\msun~\mpc^{-3}$.  Assuming a Salpeter IMF with $1-100~\msun$ and
the mass-ratio distribution $\Phi(q)=$ const, the number fraction of
massive binaries with $M_{1(2)}\geq M_{\rm crit}=25~\msun$ is $f_{\rm
  MB}=3.8\times 10^{-4}$, and the number density of PopII BBHs is
$N_{\rm MB,II}\simeq 1.5\times 10^3~\mpc^{-3}$ (for $f_{\rm
  bin}=0.7$).  The number density of PopII and PopIII binaries is
almost identical, even though the total stellar mass in PopII stars is
much higher than in PopIII stars.  For comparison, in
Fig. \ref{fig:MB_ion} we also show the ratio of the number of massive
binaries to ionizing photons for a PopII Salpeter IMF with
$1-100~\msun$ (dashed), where $\eta_{\rm ion}=8710$ is assumed
\citep{Samui_2007}.  
This shows that PopII stars produce fewer BBHs per ionizing photon.
Nevertheless, with the normalization above,
we find that PopII stars produce an ionizing background of $J_{\rm
  \nu, bg}\simeq 5\times 10^{-22}$ erg s$^{-1}$ cm$^{-2}$ Hz$^{-1}$
sr$^{-1}$ (in physical units), consistent with $\approx50\%$ of the
background measured at $z\sim 3$ \citep[e.g.][]{Steidel_2001}.
We note that 90\% of present-day stars form between $0<z<3$; however, the
more recent, more metal-rich generations at $z<3$ are not expected
to dominate the GWB, assuming a merger rate of BBHs by e.g., \cite{Domiink_2013}.

\section{Gravitational waves from PopIII binary black hole mergers}

We next consider the evolution of massive binaries to estimate the
merger rate of PopIII BBHs.  As discussed by \citetalias{K14} in the
context of their population synthesis model, most massive PopIII
binaries can evolve into BBHs because in the absence of metals (1)
mass loss from the metal-free surface is suppressed, and (2) stars are
compact, which suppresses the deleterious effects of stellar binary
interactions and mergers.  Furthermore, PopIII stars with $\sim
30~\msun$ are unlikely to evolve into red giants \citep{2001A&A...371..152M},
and as a result, the typical chirp mass of PopIII BBHs is $\langle
M_{\rm chirp} \rangle \simeq 30~\msun$ for both a flat and a Salpeter
IMF with $10-100~\msun$ (\citetalias{K14}).

By comparison, when massive PopII stars experience strong mass loss,
they go through a red giant phase.  These effects would reduce the average PopII BBH
chirp mass to $\sim 10~\msun$ \citep{Kowalska_2015}, and suppress the
formation of BBHs somewhat \citep{2010ApJ...715L.138B}.  
Furthermore, the average separation of PopII BBHs which escape mergers during the red giant phases
would be larger, and their GW inspiral time is longer, so that
only a few percent merges within a Hubble time (\citetalias{K14}).

\begin{figure}
\begin{center}
\includegraphics[width=75mm]{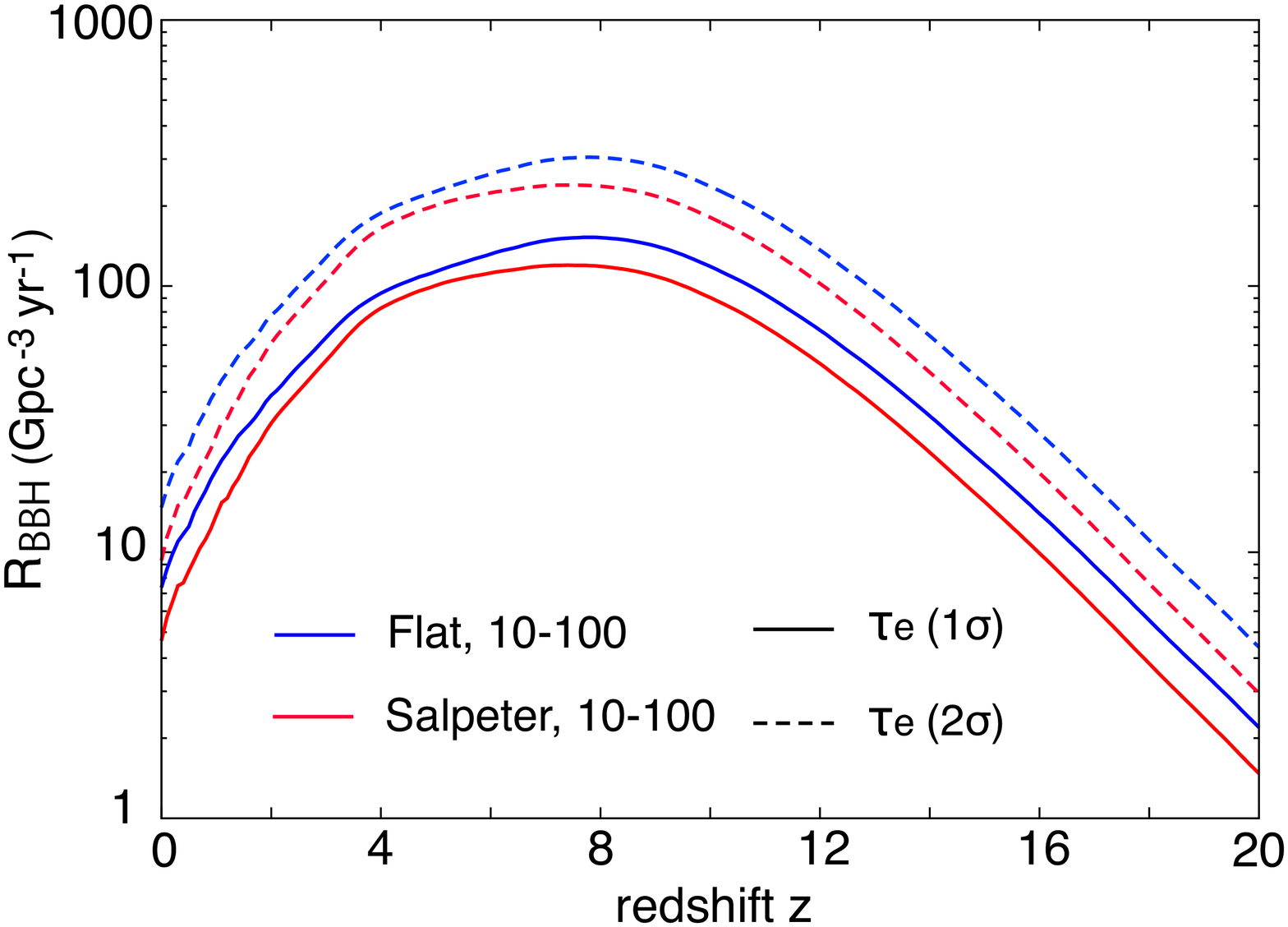}
\caption{Merger rate of PopIII BBHs for different assumed IMFs, as in
  Fig.~\ref{fig:MB_ion}.  The data is taken from \citetalias{K14}, but
  renormalized
  to be consistent with the electrons scattering optical depth
  $\tau_e$ measured by {\it Planck} within the $1\sigma$ (solid) and $2\sigma$ (dashed) error
  (Eq. \ref{eq:rho_max} with $f_{\rm esc,m}=0.1$ and $\eta_{\rm ion}=5\times 10^4$).  }
\label{fig:GW_rate}
\end{center}
\end{figure}

Fig. \ref{fig:GW_rate} shows the evolution of the PopIII BBH merger
rate $R_{\rm BBH}$ for the flat (blue) and Salpeter (red) IMF with
$10-100~\msun$.  
The distribution of the initial binary separation $a$ is assumed to be $\propto a^{-1}$ \citep{Abt_1983}.
The rates are normalized using the cumulative mass
density of PopIII stars consistent with the {\it Planck} $\tau_e$
within the $1\sigma$ (solid) and $2\sigma$ (dashed) error
(Eq. \ref{eq:rho_max} for $f_{\rm esc,m}=0.1$ and $\eta_{\rm
  ion}=5\times 10^4$), with the redshift-dependence of the SFR following \cite{DeSouza_2011}.
The rates peak between $z\approx 4-10$ at $\ga 100~\gpc^{-3}~{\rm yr}^{-1}$, 
with most merging PopIII BBHs unresolved by AdLIGO/Virgo, and with
$\ga 10^6$ PopIII BBHs contributing to a strongly redshifted GWB over five years.
The merger rates decrease toward low redshift once PopIII star
formation is quenched.
However, the rate remains as high as $R_{\rm BBH}\simeq
10~\gpc^{-3}~{\rm yr}^{-1}$ even at $z\simeq 0$, because BBHs with
suitable initial separations take a Hubble time to merge.  For a
massive $30+30~\msun$ circular BBH with an initial separation of $\sim
0.2$ AU, the GW inspiral time is 10 Gyr \citep{Peter_Mathews_1963}.
Note that the merging rate is consistent with $\sim 2-400~{\rm
  Gpc}^{-3}~{\rm yr}^{-1}$ inferred from GW150914
\citep{Abbott_2016_rate}.

We estimate the spectrum of the PopIII GWB as
\begin{equation}
\rho_{\rm c}c^2\Omega_{\rm gw}(f)=
\int _{z_{\rm min}}^\infty
\frac{R_{\rm BBH}}{1+z}\frac{dt}{dz}\left(f_r\frac{dE_{\rm gw}}{df_r}\right) \Big| _{f_r=f(1+z)}dz
\label{eq:omega_gw_2}
\end{equation}
\citep{Phinney_2001}, where $f$ and $f_r$ are the GW frequencies observed at $z=0$ and in
the source's rest frame, respectively, and $\rho_{\rm c}$ is the
critical density of the Universe.  
We set the minimum redshift to
$z_{\rm min}=0.28$, the detection horizon of AdLIGO/Virgo.  
The GW spectrum from a coalescencing BBH is given by
\begin{equation}
\frac{dE_{\rm gw}}{df_r}=\frac{(\pi G)^{2/3}M_{\rm chirp}^{5/3}}{3}\\
\begin{cases}
f_r^{-1/3}\mathscr{F_{\rm PN}}~~~~&f_r<f_1,\\[2pt]
\omega_{\rm m} f_r^{2/3}\mathscr{G_{\rm PN}}~~&f_1\leq f_r<f_2,\\[2pt]
\dfrac{\omega_{\rm r} \sigma^4 f_r^2}{[\sigma^2+4(f_r-f_2)^2]^2}
&f_2\leq f_r<f_3,\\[2pt]
\end{cases}
\label{eq:omega_gw_1}
\end{equation}
where $E_{\rm gw}$ is the energy emitted in GWs, $M_{\rm chirp}\equiv
(M_1M_2)^{3/5}/(M_1+M_2)^{1/5}$ is the chirp mass, and $f_{i}$ ($i=1,
2, 3$) and $\sigma$ are frequencies that characterize the
inspiral-merger-ringdown waveforms, $\omega_{\rm m(r)}$ are normalization constants
chosen so as to make the waveform continuous,
and the Post-Newtonian correction factors of $\mathscr{F(G)_{\rm PN}}$ \citep{Ajith_2011}.
We assume that the BBHs have circular orbits
because the PopIII BBHs should circularize by the time they move into the LIGO
band (see Fig.3 in \citealt{Abbott_2016_Astro}).
Note that the background spectrum in the inspiral phase
 scales with frequency as $\Omega_{\rm
  gw}(f)\propto f^{2/3}$.

\begin{figure}
\begin{center}
\includegraphics[width=75mm]
{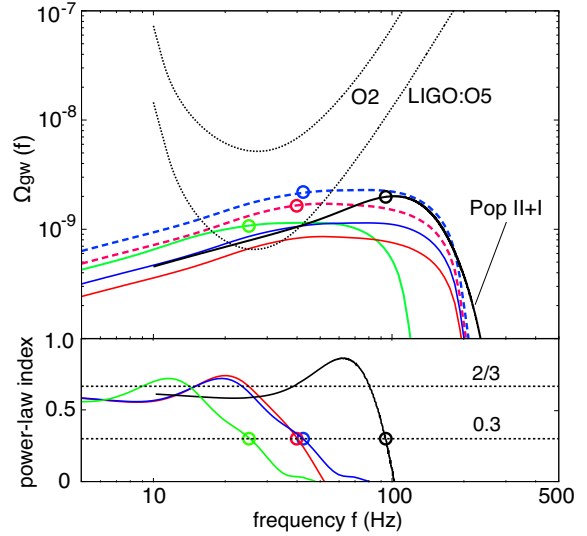}
\caption{{\it Top}: spectra of GWB produced
  by PopIII BBHs
 for the same IMFs, $f_{\rm esc,m}$ and $\tau_{\rm e}$ as in
 Fig.~\ref{fig:GW_rate} (blue and red curves).  We assume binaries
 with the average chirp mass of $\langle M_{\rm chirp}\rangle
 =30~\msun$ on circular orbits.  The background expected from all
unresolved PopII+PopI BBHs is shown for reference \citep[solid black
   curve,][their fiducial model]{LIGO_back_2016}.
Black dotted curves show the expected sensitivity of 
AdLIGO/Virgo in the observing runs O2 and O5.  
The green solid curve is the same as the blue solid curve, but with a
higher chirp mass of $\langle M_{\rm chirp}\rangle=50~\msun$ and with a lower merging rate by a factor of $3/5$. 
{\it Bottom}: the spectral index; open circles mark the frequencies above which $\alpha<0.3$.
}
\label{fig:Omega_gw}
\end{center}
\end{figure}

Fig. \ref{fig:Omega_gw} shows spectra of the GWB produced by PopIII
BBHs for the same IMFs, $f_{\rm esc,m}$ and $\tau_{\rm e}$ as in
Fig.~\ref{fig:GW_rate} (blue and red curves).  For the two IMFs with
$10-100~\msun$ and a flat mass ratio distribution, the typical merger
is an equal-mass binary with a chirp mass of $\langle M_{\rm
  chirp}\rangle \simeq 30~\msun$ (\citetalias{K14,K16}).
For comparison, we show the background produced by all PopII/I BBHs
(black solid curve), which typically merge at $z \la 2-4$
\citep{Domiink_2013,LIGO_back_2016}.
For all cases shown, the GWB
from PopIII BBHs is higher than the expected sensitivity of the
AdLIGO/Virgo detectors in the observing run O5 (black dotted curves).
The typical chirp mass and redshift of PopIII BBHs are both higher
than for PopII/I BBHs ( $\sim 30~\msun$ vs $\sim 10~\msun$ and $z\sim
8$ vs. $z\sim 3$), causing the GW frequency to be redshifted.  As a
result, the spectrum in the AdLIGO/Virgo band becomes flatter than the
canonical $\Omega_{\rm gw}(f)\propto f^{2/3}$ expected from
lower-redshift and lower-mass PopII/I sources.
\citet{Kowalska_2012} also have noted the spectral flattening by assuming 
a different chirp mass distribution and a high PopIII SFR
which is inconsistent with the Planck result and does not include 
important physics (e.g. LW feedback, metal enrichment and reionization).
In the bottom panel,
the open circles mark the frequencies above which the spectral index
falls below $0.3$; this critical frequency is $\sim~40$ Hz, well inside the
AdLIGO/Virgo band.
The deviation could be detectable with $S/N\sim 3$ in the O5 run \citep{LIGO_back_2016}.
Note that PopII/I BBHs can produce such a significant flattening of the GWB spectrum 
at $\sim 100$ Hz (black curve; see also \citealt{Kowalska_2015} that show a similar flattening
at $\ga 70-100$ Hz, depending on their models.)
Although a sub-dominant population of massive PopII BBHs with $\ga 30~\msun$
would form, depending on a model of cosmic metal enrichment \citep{Belczynski_2016_b},
the severe flatting requires the majority of such massive stars, as expected only in the PopIII model 
(although of course this remains uncertain).
Finally, we increase the chirp mass to $50~\msun$ to see its impact (green solid curve).
In this case, the critical frequency shifts to $\sim 25$ Hz, allowing the deviation of the
spectral index to be measured more easily.
The spectral shape near $30-50$ Hz would be precisely observed 
by the Einstein Telescope (ET)\footnote{http://www.et-gw.eu/etdsdocument}
with the expected sensitivity of $\Omega_{\rm gw}\approx 10^{-10}-10^{-11}$ at $f\sim 30$ Hz.


\section{Summary and Discussion}

In this paper, we show that GW background produced by PopIII
binary BHs can dominate other populations at $10-100$ Hz, even
considering tight constraints on PopIII star formation by the recent
{\it Planck} optical depth measurement from the CMB.  We also find
that the spectral index of the background from PopIII BHs becomes
significantly flatter at $\ga 30$ Hz than the canonical value of $\approx 2/3$ expected from
lower-$z$ and less-massive circular binaries. 

\begin{figure}
\begin{center}
\includegraphics[width=75mm]
{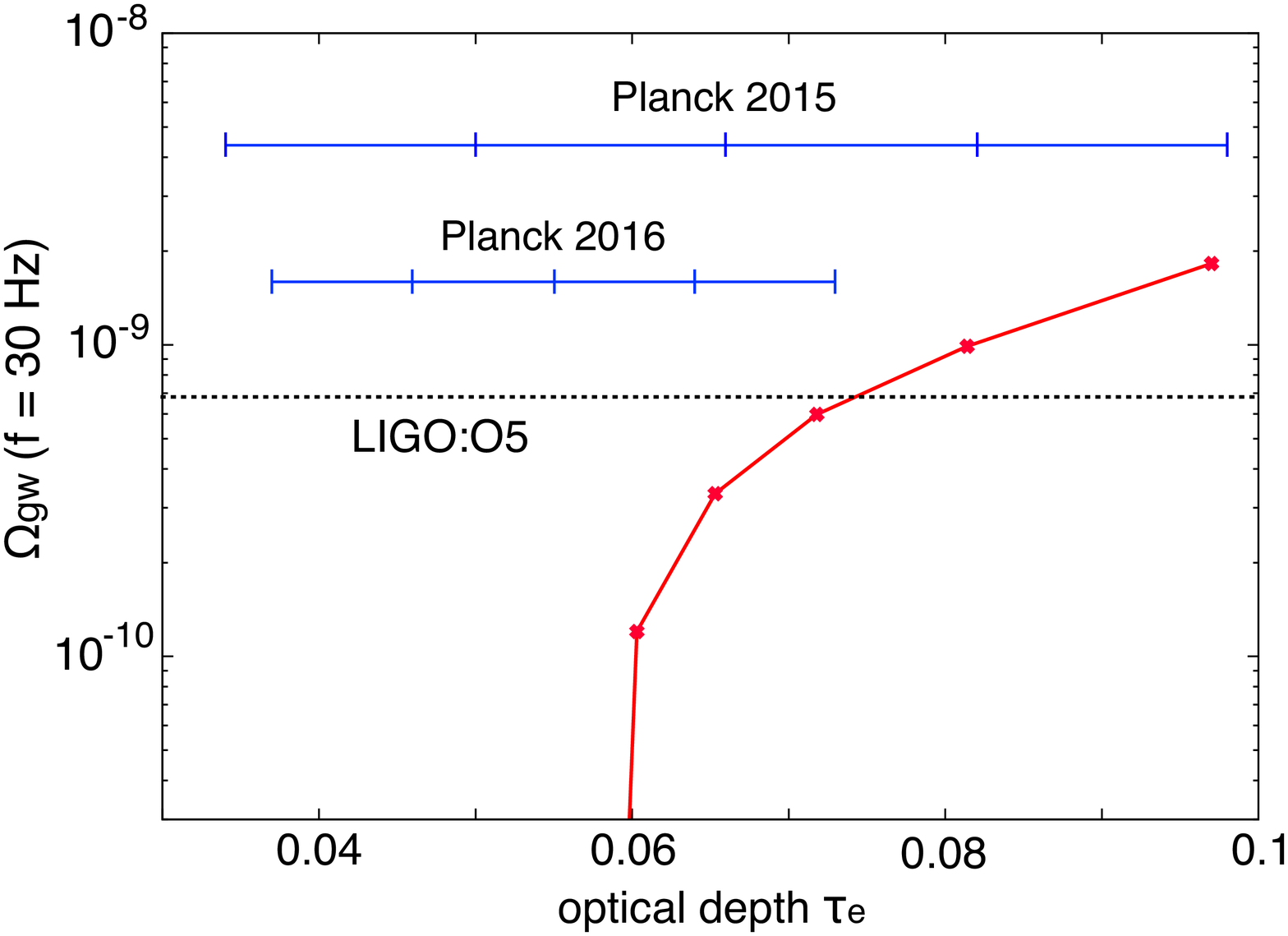}
\caption{The PopIII GWB at $f = 30$ Hz as a function of the CMB optical depth (red solid),
for $f_{\rm esc,m}=0.1$, a flat IMF with $10-100~\msun$ and $\langle M_{\rm chirp}\rangle = 30~\msun$.
For $\tau_{\rm e}\ga 0.07$, the PopIII GWB can be above the expected sensitivity of the future O5 AdLIGO/Virgo (dotted).
The results of the {\em Planck} optical depth measurement with $1\sigma$ and $2\sigma$ errors are shown
\citep{Planck_2015,Planck_2016a}.
}
\label{fig:Ogw_tau}
\end{center}
\end{figure}

As discussed in \S\ref{sec:BBH}, massive PopIII stars are also sources
of ionizing photons in the early Universe.  Thus, future observations of reionization 
and high-redshift galaxies could yield constraints on the PopIII BBH scenario.  
Fig.~\ref{fig:Ogw_tau} shows the amplitude of the PopIII GWB at $f=30$ Hz 
for different values of the CMB optical depth (red solid), 
assuming $f_{\rm esc,m}=0.1$, a flat IMF with $10-100~\msun$
and $\langle M_{\rm chirp}\rangle = 30~\msun$.
For our fiducial case, the PopIII GWB would be above 
the expected sensitivity of AdLIGO/Virgo in the observing run O5 (dotted)
as long as $\tau_{\rm e}\ga 0.07$.
However, for a small value of $\tau_{\rm e}\la0.06$\footnote{Recently, 
new values of $\tau_{\rm e}$ by {\it Planck} have been proposed as 
$0.055\pm 0.009$ and $0.058\pm 0.012$ ($1\sigma$), respectively 
\citep{Planck_2016a,Planck_2016b}.}, 
ionizing photons from PopIII stars would contribute negligibly to reionization
(e.g. \citealt{2015MNRAS.454L..76M}; \citetalias{Visbal_2015})\footnote{
A lower escape fraction of ionizing photons from PopII galaxies
($f_{\rm esc,II} < 0.1$) would allow a higher PopIII star (BBH) formation rate,
however reducing this escape fraction prevents the completion of
reionization by $z=6$.}.
A detection of the PopIII GWB by the future observations 
(O5 AdLIGO/Virgo, KAGRA\footnote{http://gwcenter.icrr.u-tokyo.ac.jp/en/} and ET), 
if any, implies that the IMF of PopIII stars would be more top-heavy, where PopIII stars 
form more BBHs but produce less ionizing photons (see Fig. \ref{fig:MB_ion}). 
The GWB (even an upper limit) would provide a
useful constraint on the PopIII BBH scenario,
combining with theoretical studies of high-$z$ star formation.

The origin of massive BBHs such as GW150914 is still uncertain.
Future detections of multiple GW events with a BBH coalescence rate of
$R_{\rm BBH}\sim 10~\gpc^{-3}~{\rm yr}^{-1}$ and a typical chirp mass
of $\sim 30~\msun$ would suggest a PopIII BBH scenario. In this case,
the GWB could also be detectable, along with a significant flattening of the spectral index.  
There are three other effects which can cause a deviation of the spectral slope from the
canonical value for circular, GW-driven, non-spinning inspirals:
(i) environmental
effects (interaction of the BHs with nearby gas and/or stars), 
(ii) high orbital eccentricities and (iii) high BH spins.  
The first two of these effects would generically steepen the spectral slope
because environmental effects remove energy
preferentially at large separations, and eccentricities move power
from low to high frequencies.  
In principle, strong BH spins anti-aligned with the orbital angular momentum
can flatten the spectrum significantly, but only at high frequencies near the
merger \citep[e.g.][]{2011ApJ...739...86Z}.
Therefore, the detection of a flattening of the spectral index of
GWB at frequencies as low as $30$ Hz would be an unique smoking gun
of a high-chirp mass, high-redshift BBH population, as
expected from PopIII stars.

\section*{Acknowledgements}
We thank Imre Bartos, Bence Kocsis, Tomoya Kinugawa, Tania Regimbau, 
David Schiminovich and Alberto Sesana for useful discussions.
This work is partially supported by the Simons Foundation through 
the Simons Society of Fellows (KI), 
by NASA through an Einstein Postdoctoral Fellowship (KK),
by the Columbia Prize Postdoctoral Fellowship in the Natural Sciences (EV)
and by NASA grants NNX11AE05G and NNX15AB19G (ZH).

{\small
\bibliography{ref}

\begin{thebibliography}{53}
\expandafter\ifx\csname natexlab\endcsname\relax\def\natexlab#1{#1}\fi

\bibitem[{{Abbott} {et~al}\mbox{.}(2016{\natexlab{a}}){Abbott}, {Abbott},
  {Abbott}, {Abernathy}, {Acernese}, {Ackley}, {Adams}, {Adams}, {Addesso},
  {Adhikari}, \& et~al.}]{Abbott_2016_Astro}
{Abbott} B.~P. {et~al.}, 2016{\natexlab{a}}, \apjl, 818, L22

\bibitem[{{Abbott} {et~al}\mbox{.}(2016{\natexlab{b}}){Abbott}, {Abbott},
  {Abbott}, {Abernathy}, {Acernese}, {Ackley}, {Adams}, {Adams}, {Addesso},
  {Adhikari}, \& et~al.}]{LIGO_back_2016}
{Abbott} B.~P. {et~al.}, 2016{\natexlab{b}}, arXiv:1602.03847

\bibitem[{{Abbott} {et~al}\mbox{.}(2016{\natexlab{c}}){Abbott}, {Abbott},
  {Abbott}, {Abernathy}, {Acernese}, {Ackley}, {Adams}, {Adams}, {Addesso},
  {Adhikari}, \& et~al.}]{Abbott_PRL_2016}
{Abbott} B.~P. {et~al.}, 2016{\natexlab{c}}, Physical Review Letters, 116,
  061102

\bibitem[{{Abbott} {et~al}\mbox{.}(2016{\natexlab{d}}){Abbott}, {Abbott},
  {Abbott}, {Abernathy}, {Acernese}, {Ackley}, {Adams}, {Adams}, {Addesso},
  {Adhikari}, \& et~al.}]{Abbott_2016_rate}
{Abbott} B.~P. {et~al.}, 2016{\natexlab{d}}, arXiv:1602.03842

\bibitem[{{Abt}(1983)}]{Abt_1983}
{Abt} H.~A., 1983, \araa, 21, 343

\bibitem[{{Adam} {et~al}\mbox{.}(2016){Adam}, {Aghanim}, {Ashdown}, {Aumont},
  {Baccigalupi}, {Ballardini}, {Banday}, {Barreiro}, {Bartolo}, {Basak},
  {Battye}, {Benabed}, {Bernard}, {Bersanelli}, {Bielewicz}, {Bock}, {Bonaldi},
  {Bonavera}, {Bond}, {Borrill}, {Bouchet}, {Bucher}, {Burigana}, {Calabrese},
  {Cardoso}, {Carron}, {Chiang}, {Colombo}, {Combet}, {Comis}, {Coulais},
  {Crill}, {Curto}, {Cuttaia}, {Davis}, {de Bernardis}, {de Rosa}, {de Zotti},
  {Delabrouille}, {Di Valentino}, {Dickinson}, {Diego}, {Dor{\'e}}, {Douspis},
  {Ducout}, {Dupac}, {Elsner}, {En{\ss}lin}, {Eriksen}, {Falgarone}, {Fantaye},
  {Finelli}, {Forastieri}, {Frailis}, {Fraisse}, {Franceschi}, {Frolov},
  {Galeotta}, {Galli}, {Ganga}, {G{\'e}nova-Santos}, {Gerbino}, {Ghosh},
  {Gonz{\'a}lez-Nuevo}, {G{\'o}rski}, {Gruppuso}, {Gudmundsson}, {Hansen},
  {Helou}, {Henrot-Versill{\'e}}, {Herranz}, {Hivon}, {Huang}, {Ili}, {Jaffe},
  {Jones}, {Keih{\"a}nen}, {Keskitalo}, {Kisner}, {Knox}, {Krachmalnicoff},
  {Kunz}, {Kurki-Suonio}, {Lagache}, {L{\"a}hteenm{\"a}ki}, {Lamarre},
  {Langer}, {Lasenby}, {Lattanzi}, {Lawrence}, {Le Jeune}, {Levrier}, {Lewis},
  {Liguori}, {Lilje}, {L{\'o}pez-Caniego}, {Ma}, {Mac{\'{\i}}as-P{\'e}rez},
  {Maggio}, {Mangilli}, {Maris}, {Martin}, {Mart{\'{\i}}nez-Gonz{\'a}lez},
  {Matarrese}, {Mauri}, {McEwen}, {Meinhold}, {Melchiorri}, {Mennella},
  {Migliaccio}, {Miville-Desch{\^e}nes}, {Molinari}, {Moneti}, {Montier},
  {Morgante}, {Moss}, {Naselsky}, {Natoli}, {Oxborrow}, {Pagano}, {Paoletti},
  {Partridge}, {Patanchon}, {Patrizii}, {Perdereau}, {Perotto}, {Pettorino},
  {Piacentini}, {Plaszczynski}, {Polastri}, {Polenta}, {Puget}, {Rachen},
  {Racine}, {Reinecke}, {Remazeilles}, {Renzi}, {Rocha}, {Rossetti}, {Roudier},
  {Rubi{\~n}o-Mart{\'{\i}}n}, {Ruiz-Granados}, {Salvati}, {Sandri},
  {Savelainen}, {Scott}, {Sirri}, {Sunyaev}, {Suur-Uski}, {Tauber}, {Tenti},
  {Toffolatti}, {Tomasi}, {Tristram}, {Trombetti}, {Valiviita}, {Van Tent},
  {Vielva}, {Villa}, {Vittorio}, {Wandelt}, {Wehus}, {White}, {Zacchei}, \&
  {Zonca}}]{Planck_2016a}
{Adam} R. {et~al.}, 2016, arXiv:1605.03507

\bibitem[{{Ade} {et~al}\mbox{.}(2015){Ade}, {Aghanim}, {Arnaud}, {Ashdown},
  {Aumont}, {Baccigalupi}, {Banday}, {Barreiro}, {Bartlett}, \&
  et~al.}]{Planck_2015}
{Ade} P.~A.~R. {et~al.}, 2015, arXiv:1502.01589

\bibitem[{{Aghanim} {et~al}\mbox{.}(2016){Aghanim}, {Ashdown}, {Aumont},
  {Baccigalupi}, {Ballardini}, {Banday}, {Barreiro}, {Bartolo}, {Basak},
  {Battye}, {Benabed}, {Bernard}, {Bersanelli}, {Bielewicz}, {Bock}, {Bonaldi},
  {Bonavera}, {Bond}, {Borrill}, {Bouchet}, {Boulanger}, {Bucher}, {Burigana},
  {Butler}, {Calabrese}, {Cardoso}, {Carron}, {Challinor}, {Chiang}, {Colombo},
  {Combet}, {Comis}, {Coulais}, {Crill}, {Curto}, {Cuttaia}, {Davis}, {de
  Bernardis}, {de Rosa}, {de Zotti}, {Delabrouille}, {Delouis}, {Di Valentino},
  {Dickinson}, {Diego}, {Dor{\'e}}, {Douspis}, {Ducout}, {Dupac}, {Efstathiou},
  {Elsner}, {En{\ss}lin}, {Eriksen}, {Falgarone}, {Fantaye}, {Finelli},
  {Forastieri}, {Frailis}, {Fraisse}, {Franceschi}, {Frolov}, {Galeotta},
  {Galli}, {Ganga}, {G{\'e}nova-Santos}, {Gerbino}, {Ghosh},
  {Gonz{\'a}lez-Nuevo}, {G{\'o}rski}, {Gratton}, {Gruppuso}, {Gudmundsson},
  {Hansen}, {Helou}, {Henrot-Versill{\'e}}, {Herranz}, {Hivon}, {Huang},
  {Ilic}, {Jaffe}, {Jones}, {Keih{\"a}nen}, {Keskitalo}, {Kisner}, {Knox},
  {Krachmalnicoff}, {Kunz}, {Kurki-Suonio}, {Lagache}, {Lamarre}, {Langer},
  {Lasenby}, {Lattanzi}, {Lawrence}, {Le Jeune}, {Leahy}, {Levrier}, {Liguori},
  {Lilje}, {L{\'o}pez-Caniego}, {Ma}, {Mac{\'{\i}}as-P{\'e}rez}, {Maggio},
  {Mangilli}, {Maris}, {Martin}, {Mart{\'{\i}}nez-Gonz{\'a}lez}, {Matarrese},
  {Mauri}, {McEwen}, {Meinhold}, {Melchiorri}, {Mennella}, {Migliaccio},
  {Miville-Desch{\^e}nes}, {Molinari}, {Moneti}, {Montier}, {Morgante}, {Moss},
  {Mottet}, {Naselsky}, {Natoli}, {Oxborrow}, {Pagano}, {Paoletti},
  {Partridge}, {Patanchon}, {Patrizii}, {Perdereau}, {Perotto}, {Pettorino},
  {Piacentini}, {Plaszczynski}, {Polastri}, {Polenta}, {Puget}, {Rachen},
  {Racine}, {Reinecke}, {Remazeilles}, {Renzi}, {Rocha}, {Rossetti}, {Roudier},
  {Rubi{\~n}o-Mart{\'{\i}}n}, {Ruiz-Granados}, {Salvati}, {Sandri},
  {Savelainen}, {Scott}, {Sirri}, {Sunyaev}, {Suur-Uski}, {Tauber}, {Tenti},
  {Toffolatti}, {Tomasi}, {Tristram}, {Trombetti}, {Valiviita}, {Van Tent},
  {Vibert}, {Vielva}, {Villa}, {Vittorio}, {Wandelt}, {Watson}, {Wehus},
  {White}, {Zacchei}, \& {Zonca}}]{Planck_2016b}
{Aghanim} N. {et~al.}, 2016, arXiv:1605.02985

\bibitem[{{Ahn} {et~al}\mbox{.}(2012){Ahn}, {Iliev}, {Shapiro}, {Mellema},
  {Koda}, \& {Mao}}]{Ahn_2012}
{Ahn} K., {Iliev} I.~T., {Shapiro} P.~R., {Mellema} G., {Koda} J., {Mao} Y.,
  2012, \apjl, 756, L16

\bibitem[{{Ajith} {et~al}\mbox{.}(2011){Ajith}, {Hannam}, {Husa}, {Chen},
  {Br{\"u}gmann}, {Dorband}, {M{\"u}ller}, {Ohme}, {Pollney}, {Reisswig},
  {Santamar{\'{\i}}a}, \& {Seiler}}]{Ajith_2011}
{Ajith} P. {et~al.}, 2011, Physical Review Letters, 106, 241101

\bibitem[{{Bartos} {et~al}\mbox{.}(2016){Bartos}, {Kocsis}, {Haiman}, \&
  {M{\'a}rka}}]{Bartos_2016}
{Bartos} I., {Kocsis} B., {Haiman} Z., {M{\'a}rka} S., 2016, arXiv:1602.03831

\bibitem[{{Belczynski}, {Bulik} \& {Rudak}(2004){Belczynski}, {Bulik}, \&
  {Rudak}}]{Belczynski_2004}
{Belczynski} K., {Bulik} T., {Rudak} B., 2004, \apjl, 608, L45

\bibitem[{{Belczynski} {et~al}\mbox{.}(2010){Belczynski}, {Dominik}, {Bulik},
  {O'Shaughnessy}, {Fryer}, \& {Holz}}]{2010ApJ...715L.138B}
{Belczynski} K., {Dominik} M., {Bulik} T., {O'Shaughnessy} R., {Fryer} C.,
  {Holz} D.~E., 2010, \apjl, 715, L138

\bibitem[{{Belczynski} {et~al}\mbox{.}(2016){Belczynski}, {Holz}, {Bulik}, \&
  {O'Shaughnessy}}]{Belczynski_2016_b}
{Belczynski} K., {Holz} D.~E., {Bulik} T., {O'Shaughnessy} R., 2016,
  arXiv:1602.04531

\bibitem[{{Bond} \& {Carr}(1984)}]{1984MNRAS.207..585B}
{Bond} J.~R., {Carr} B.~J., 1984, \mnras, 207, 585

\bibitem[{{Bromm} \& {Yoshida}(2011)}]{2011ARA&A..49..373B}
{Bromm} V., {Yoshida} N., 2011, \araa, 49, 373

\bibitem[{{de Souza}, {Yoshida} \& {Ioka}(2011){de Souza}, {Yoshida}, \&
  {Ioka}}]{DeSouza_2011}
{de Souza} R.~S., {Yoshida} N., {Ioka} K., 2011, \aap, 533, A32

\bibitem[{{Dominik} {et~al}\mbox{.}(2012){Dominik}, {Belczynski}, {Fryer},
  {Holz}, {Berti}, {Bulik}, {Mandel}, \& {O'Shaughnessy}}]{Dominik_2012}
{Dominik} M., {Belczynski} K., {Fryer} C., {Holz} D.~E., {Berti} E., {Bulik}
  T., {Mandel} I., {O'Shaughnessy} R., 2012, \apj, 759, 52

\bibitem[{{Dominik} {et~al}\mbox{.}(2013){Dominik}, {Belczynski}, {Fryer},
  {Holz}, {Berti}, {Bulik}, {Mandel}, \& {O'Shaughnessy}}]{Domiink_2013}
{Dominik} M., {Belczynski} K., {Fryer} C., {Holz} D.~E., {Berti} E., {Bulik}
  T., {Mandel} I., {O'Shaughnessy} R., 2013, \apj, 779, 72

\bibitem[{{Haiman}, {Abel} \& {Rees}(2000){Haiman}, {Abel}, \&
  {Rees}}]{Haiman_2000}
{Haiman} Z., {Abel} T., {Rees} M.~J., 2000, \apj, 534, 11

\bibitem[{{Haiman}, {Rees} \& {Loeb}(1997){Haiman}, {Rees}, \&
  {Loeb}}]{1997ApJ...476..458H}
{Haiman} Z., {Rees} M.~J., {Loeb} A., 1997, \apj, 476, 458

\bibitem[{{Hartwig} {et~al}\mbox{.}(2016){Hartwig}, {Volonteri}, {Bromm},
  {Klessen}, {Barausse}, {Magg}, \& {Stacy}}]{Hartwig_2016}
{Hartwig} T., {Volonteri} M., {Bromm} V., {Klessen} R.~S., {Barausse} E.,
  {Magg} M., {Stacy} A., 2016, arXiv:1603.05655

\bibitem[{{Heger} \& {Woosley}(2002)}]{2002ApJ...567..532H}
{Heger} A., {Woosley} S.~E., 2002, \apj, 567, 532

\bibitem[{{Hirano} {et~al}\mbox{.}(2014){Hirano}, {Hosokawa}, {Yoshida},
  {Umeda}, {Omukai}, {Chiaki}, \& {Yorke}}]{2014ApJ...781...60H}
{Hirano} S., {Hosokawa} T., {Yoshida} N., {Umeda} H., {Omukai} K., {Chiaki} G.,
  {Yorke} H.~W., 2014, \apj, 781, 60

\bibitem[{{Kinugawa} {et~al}\mbox{.}(2014){Kinugawa}, {Inayoshi}, {Hotokezaka},
  {Nakauchi}, \& {Nakamura}}]{K14}
{Kinugawa} T., {Inayoshi} K., {Hotokezaka} K., {Nakauchi} D., {Nakamura} T.,
  2014, \mnras, 442, 2963

\bibitem[{{Kinugawa} {et~al}\mbox{.}(2016){Kinugawa}, {Miyamoto}, {Kanda}, \&
  {Nakamura}}]{K16}
{Kinugawa} T., {Miyamoto} A., {Kanda} N., {Nakamura} T., 2016, \mnras, 456,
  1093

\bibitem[{{Kitayama} {et~al}\mbox{.}(2004){Kitayama}, {Yoshida}, {Susa}, \&
  {Umemura}}]{2004ApJ...613..631K}
{Kitayama} T., {Yoshida} N., {Susa} H., {Umemura} M., 2004, \apj, 613, 631

\bibitem[{{Komatsu} {et~al}\mbox{.}(2011){Komatsu}, {Smith}, {Dunkley},
  {Bennett}, {Gold}, {Hinshaw}, {Jarosik}, {Larson}, {Nolta}, {Page},
  {Spergel}, {Halpern}, {Hill}, {Kogut}, {Limon}, {Meyer}, {Odegard}, {Tucker},
  {Weiland}, {Wollack}, \& {Wright}}]{Komatsu_2011}
{Komatsu} E. {et~al.}, 2011, \apjs, 192, 18

\bibitem[{{Kowalska}, {Bulik} \& {Belczynski}(2012){Kowalska}, {Bulik}, \&
  {Belczynski}}]{Kowalska_2012}
{Kowalska} I., {Bulik} T., {Belczynski} K., 2012, \aap, 541, A120

\bibitem[{{Kowalska-Leszczynska} {et~al}\mbox{.}(2015){Kowalska-Leszczynska},
  {Regimbau}, {Bulik}, {Dominik}, \& {Belczynski}}]{Kowalska_2015}
{Kowalska-Leszczynska} I., {Regimbau} T., {Bulik} T., {Dominik} M.,
  {Belczynski} K., 2015, \aap, 574, A58

\bibitem[{{Kulczycki} {et~al}\mbox{.}(2006){Kulczycki}, {Bulik},
  {Belczy{\'n}ski}, \& {Rudak}}]{2006A&A...459.1001K}
{Kulczycki} K., {Bulik} T., {Belczy{\'n}ski} K., {Rudak} B., 2006, \aap, 459,
  1001

\bibitem[{{Loeb}(2016)}]{Loeb_2016}
{Loeb} A., 2016, \apjl, 819, L21

\bibitem[{{Mandel} \& {de Mink}(2016)}]{Mandel_2016}
{Mandel} I., {de Mink} S.~E., 2016, \mnras

\bibitem[{{Marigo} {et~al}\mbox{.}(2001){Marigo}, {Girardi}, {Chiosi}, \&
  {Wood}}]{2001A&A...371..152M}
{Marigo} P., {Girardi} L., {Chiosi} C., {Wood} P.~R., 2001, \aap, 371, 152

\bibitem[{{Mitra}, {Choudhury} \& {Ferrara}(2015){Mitra}, {Choudhury}, \&
  {Ferrara}}]{2015MNRAS.454L..76M}
{Mitra} S., {Choudhury} T.~R., {Ferrara} A., 2015, \mnras, 454, L76

\bibitem[{{O'Leary}, {Meiron} \& {Kocsis}(2016){O'Leary}, {Meiron}, \&
  {Kocsis}}]{OLeary_2016}
{O'Leary} R.~M., {Meiron} Y., {Kocsis} B., 2016, arXiv:1602.02809

\bibitem[{{Omukai} {et~al}\mbox{.}(2005){Omukai}, {Tsuribe}, {Schneider}, \&
  {Ferrara}}]{2005ApJ...626..627O}
{Omukai} K., {Tsuribe} T., {Schneider} R., {Ferrara} A., 2005, \apj, 626, 627

\bibitem[{{P{\'e}rez-Gonz{\'a}lez}
  {et~al}\mbox{.}(2008){P{\'e}rez-Gonz{\'a}lez}, {Rieke}, {Villar}, {Barro},
  {Blaylock}, {Egami}, {Gallego}, {Gil de Paz}, {Pascual}, {Zamorano}, \&
  {Donley}}]{2008ApJ...675..234P}
{P{\'e}rez-Gonz{\'a}lez} P.~G. {et~al.}, 2008, \apj, 675, 234

\bibitem[{{Peters} \& {Mathews}(1963)}]{Peter_Mathews_1963}
{Peters} P.~C., {Mathews} J., 1963, Physical Review, 131, 435

\bibitem[{{Phinney}(2001)}]{Phinney_2001}
{Phinney} E.~S., 2001, arXiv:0108028

\bibitem[{{Portegies Zwart} \& {McMillan}(2000)}]{PortegiesZwart_2000}
{Portegies Zwart} S.~F., {McMillan} S.~L.~W., 2000, \apjl, 528, L17

\bibitem[{{Ricotti}, {Gnedin} \& {Shull}(2002){Ricotti}, {Gnedin}, \&
  {Shull}}]{Ricotti_2002}
{Ricotti} M., {Gnedin} N.~Y., {Shull} J.~M., 2002, \apj, 575, 49

\bibitem[{{Samui}, {Srianand} \& {Subramanian}(2007){Samui}, {Srianand}, \&
  {Subramanian}}]{Samui_2007}
{Samui} S., {Srianand} R., {Subramanian} K., 2007, \mnras, 377, 285

\bibitem[{{Sana} {et~al}\mbox{.}(2012){Sana}, {de Mink}, {de Koter}, {Langer},
  {Evans}, {Gieles}, {Gosset}, {Izzard}, {Le Bouquin}, \&
  {Schneider}}]{Sana_2012}
{Sana} H. {et~al.}, 2012, Science, 337, 444

\bibitem[{{Schaerer}(2002)}]{2002A&A...382...28S}
{Schaerer} D., 2002, \aap, 382, 28

\bibitem[{{Sobacchi} \& {Mesinger}(2013)}]{SobacchiMesinger_2013}
{Sobacchi} E., {Mesinger} A., 2013, \mnras, 432, 3340

\bibitem[{{Stacy} \& {Bromm}(2013)}]{2013MNRAS.433.1094S}
{Stacy} A., {Bromm} V., 2013, \mnras, 433, 1094

\bibitem[{{Steidel}, {Pettini} \& {Adelberger}(2001){Steidel}, {Pettini}, \&
  {Adelberger}}]{Steidel_2001}
{Steidel} C.~C., {Pettini} M., {Adelberger} K.~L., 2001, \apj, 546, 665

\bibitem[{{Stone}, {Metzger} \& {Haiman}(2016){Stone}, {Metzger}, \&
  {Haiman}}]{Stone_2016}
{Stone} N.~C., {Metzger} B.~D., {Haiman} Z., 2016, arXiv:1602.04226

\bibitem[{{Visbal}, {Haiman} \& {Bryan}(2015){Visbal}, {Haiman}, \&
  {Bryan}}]{Visbal_2015}
{Visbal} E., {Haiman} Z., {Bryan} G.~L., 2015, \mnras, 453, 4456

\bibitem[{{Vogelsberger} {et~al}\mbox{.}(2013){Vogelsberger}, {Genel},
  {Sijacki}, {Torrey}, {Springel}, \& {Hernquist}}]{2013MNRAS.436.3031V}
{Vogelsberger} M., {Genel} S., {Sijacki} D., {Torrey} P., {Springel} V.,
  {Hernquist} L., 2013, \mnras, 436, 3031

\bibitem[{{Wise} {et~al}\mbox{.}(2014){Wise}, {Demchenko}, {Halicek}, {Norman},
  {Turk}, {Abel}, \& {Smith}}]{Wise_2014}
{Wise} J.~H., {Demchenko} V.~G., {Halicek} M.~T., {Norman} M.~L., {Turk} M.~J.,
  {Abel} T., {Smith} B.~D., 2014, \mnras, 442, 2560

\bibitem[{{Zhu} {et~al}\mbox{.}(2011){Zhu}, {Howell}, {Regimbau}, {Blair}, \&
  {Zhu}}]{2011ApJ...739...86Z}
{Zhu} X.-J., {Howell} E., {Regimbau} T., {Blair} D., {Zhu} Z.-H., 2011, \apj,
  739, 86

\end{thebibliography}
}
\end{document}